\documentclass[12pt,showpacs,preprintnumbers,amsmath,amssymb,nofootinbib,superscriptaddress]{article}
\usepackage{graphicx}
\usepackage{dcolumn}
\usepackage{bm}
\usepackage{graphics}
\usepackage{amssymb}
\usepackage{amscd}
\usepackage{afterpage}
\usepackage{float,times}
\usepackage{subfigure}
\usepackage{rotating}
\usepackage{multirow}
\usepackage{epsfig}
\usepackage{theorem}
\usepackage{moreverb}
\usepackage{euscript}
\usepackage{psfrag}

\begin{document}
{\hbox to\hsize{\hfill August 2011 }}

\bigskip \vspace{3\baselineskip}

\begin{center}
{\bf \Large 
Once more: gravity is not an entropic force }

\bigskip

\bigskip

{\bf Archil Kobakhidze \\}

\smallskip

{ \small \it
ARC Center of Excellence for Particle Physics at the Terascale, \\
School of Physics, The University of Melbourne, Victoria 3010, Australia \\
 archilk@unimelb.edu.au
\\}

\bigskip
 
\bigskip

{\large \bf Abstract}

\end{center}
\noindent 
{\small 
We argue that neutron interference experiments and experiments on gravitational bound states of neutron unambiguously disprove entropic origin of gravitation. The criticism expressed in a recent paper \cite{Chaichian:2011xc} concerning our arguments against entropic gravity is shown to be invalid.}


\paragraph{Verlinde vs Newton.} E. Verlinde's idea \cite{Verlinde:2010hp} on the entropic origin of gravitation has recently attracted considerable attention. Although  certain aspects  of  Ref.  \cite{Verlinde:2010hp} strongly reassemble earlier works \cite{Jacobson:1995ab}, \cite{Padmanabhan:2009vy}, there is a crucial new physics input in \cite{Verlinde:2010hp}  that makes Verlinde's proposal truly unorthodox approach to the theory of gravitation. Namely, the gravitational force in Verlinde's approach is associated with an entropic force. 

Back in the 19th century, the overwhelming success of Newtonian dynamics, and in particular, of Newtonian gravitational and Coulomb electrostatic force laws, had led to an important insight that the fundamental forces ($\vec F$) of nature could be modelled using potentials ($\Phi(\vec x)$) which satisfy Laplace's equation: 
\begin{equation}
\vec F=-m\vec \triangledown \Phi~,~~ \vec \triangledown \times \vec F=0~\Longrightarrow \triangle \Phi =0~, 
\label{1}
\end{equation}
where, $\vec \triangledown =\left(\frac{ \partial}{\partial x_1}, \frac{ \partial}{\partial x_2}, \frac{ \partial}{\partial x_3} \right)$ and $\triangle= \vec \triangledown \cdot \vec \triangledown$ 
Further relativistic generalization of the notion of non-relativistic potentials resulted in classical field theories, such as Maxwell's electrodynamics and Einstein's General Relativity. 
Quantum theory has brought necessity for quantizing these fields and introduced the notion of interaction-mediating particles, such as photon and graviton. 
These theories correctly describe experiments within the  range of their applicability. 

Despite of these remarkable successes researchers are tempted to explore alternative ways to describe known interactions, most notably the gravitational interactions. One such alternative way has been proposed by E. Verlinde \cite{Verlinde:2010hp} in the early 2010.  According to his proposal the gravitational force is not a fundamental force, but the force that emerges at large distance scales from some (yet to be specified) quantum reality that is tied with the holographic ideas. That is, there is no classical Newtonian potential and hence no quantum field generalization of thereof, no gravitons, etc.  The gravitational force in Verlinde's theory is described by the gradient of an entropy ($S$) associated with the position of gravitating bodies. Thus, instead of (\ref{1}), now we have:
 \begin{equation}
\vec F=T \vec \triangledown S~,  
\label{2}
\end{equation}  
The above equation describes Newtonian gravitational force acting on a test particle of mass $m$ by a particle of mass $M$ if the following identifications are assumed. The temperature $T$ in Eq. (\ref{2}) is associated with the Unruh temperature, 
\begin{equation}
T=\frac{g}{2\pi}~,
\label{3}
\end{equation}
where $g=G_N M/r^2$ is the free fall acceleration [$G_N$ is the Newton constant and $r$ is a distance between gravitating particles], while the gradient of entropy associated with the position $\vec r=(x_1, x_2, x_3)$ of particle $m$ relative to particle $M$ is given by \cite{Verlinde:2010hp}:
\begin{equation}
\vec \triangledown S= 2\pi m \frac{\vec r}{r}~.
\label{4}
\end{equation}
The origin of Eq. (\ref{4}) is not fully determined in Verlinde's work \cite{Verlinde:2010hp}.  It has been assumed that the underlying physics is given by an ensemble of microstates leaving on a holographic screen. This holographic screen contains maximum entropy that can be fitted in a given volume surrounded by the screen. The space in this approach is also emergent phenomenon, therefore $\vec r$ is just a macroscopic parameter that describes an ensemble of microstates. The entropy $S$ associated with this ensemble depends  on macroscopic parameters $\vec r$ and energy $E(\vec r)$, $S=S(\vec r, E)$ such that:
\begin{equation}
\frac{dS}{dx_i}\equiv \triangledown_i S+\frac{\partial S}{\partial E}\triangledown_i E=0~,
\label{5}
\end{equation}
Hence, $\vec \triangledown S=\frac{1}{T}\vec \triangledown E$. This suggests that up to an inessential additive constant $E=-m\Phi$. However, one must remember that the Newtonian potential in Verlinde's approach is merely a macroscopic parameter that characterizes the ensemble of microstates.

To summarize, according to Verlinde, gravitational force is a \emph{macroscopic} phenomenon between material bodies caused by the statistical tendency of a system  to increase its entropy  as the relative position of gravitating bodies changes (a change of matter distribution in the gravitating system).  We would like to stress that this is fundamentally different from the standard Newtonian approach, which is based on a local potential, and thus predicts existence of universal force of gravitation also for an arbitrary microscopic system\footnote{There is somewhat misleading claim in the literature \cite{Hossenfelder:2010ih} that Verlinde's and Newton's approach to gravity are actually equivalent. However, the formal mathematical equivalence established in \cite{Hossenfelder:2010ih} by inverting the chain of Verlinde's equations lacks appropriate physical considerations. E.g.,  `entropy' and `temperature'  defined respectively in Eqs. (1) and (2) of \cite{Hossenfelder:2010ih} lack physical meaning and represent just mathematical symbols.}. 

In Ref. \cite{Kobakhidze:2010mn} we have shown that, unlike the Newtonian theory, Verlinde's gravity incorrectly describes quantum-mechanical systems and thus must be dismissed. This conclusion is in full accord with the earlier unpublished article \cite{motl}. Recently,  there appear some criticism of my work \cite{Kobakhidze:2010mn}  in \cite{Chaichian:2011xc}, where the authors reached the opposite conclusion. The purpose of this paper is to restate that quantum mechanical experiments on gravitational bound states of neutron do \emph{unambiguously} disprove Verlinde's approach to gravitation and to show that criticism in \cite{Chaichian:2011xc} is invalid. In addition, we also show explicitly that Verlinde's theory disagree with the neutron interference experiments \cite{Colella:1975dq} as it was originally  suggested in \cite{motl}.   

\paragraph{Quantm mechanics with Verlinde's gravity.} 

The reason why Verlinde's and Newton's theories of gravitation give different results  for microscopic systems is the following. An \emph{inevitable} consequence of Verlinde's approach is that a test particle $m$ is described by a statistically large number $n(r)$ of microstates which depends on the position of the particle with respect to another particle $M$:
\begin{equation}
n(r)=\frac{2m}{T}=\frac{4\pi r^2}{G_N}\frac{m}{M}~
\label{}
\end{equation} 
(see Eq. (3.14) in \cite{Verlinde:2010hp} ). Hence the test particle carries the entropy $S_{\cal N}(r)$ which changes with the distance $r$ as:
\begin{eqnarray}
\Delta S_{\cal N}=\Delta \log\left(\frac{m}{M}N(r)\right)=\Delta \log N(r)=\Delta S \nonumber \\ 
=2\pi m \Delta r~,
\label{p0}
\end{eqnarray}
where in the last step we have used Eq. (3.6) from \cite{Verlinde:2010hp}. While this approach correctly describes gravitation between macroscopic particles it fails to do so for microscopic (quantum-mechanical) particles.  Indeed, position and momentum in classical mechanics are well-defined quantities that unambiguously characterize given state of a system. In quantum mechanics, however,  they are represented by operators which do not commute with each other. In position (coordinate) representation, the momentum operator generates spatial translations. Hence, if Verlinde's theory is correct, an action of this operator on a particle state must lead to a change in entropy. Obviously, such an operator would be non-Hermitian. In fact, we have shown in \cite{Kobakhidze:2010mn} that the momentum operator is:
\begin{equation}
\hat p=-i\frac{\partial}{\partial r} -i2\pi m~.
\label{p1}
\end{equation}
Therefore, the Hamiltonian of a quantum-mechanical particle moving in the classical gravitational field produced by another particle $M$ is given as:
\begin{equation}
\hat H \equiv \frac{\hat p^2}{2m}+V_{\rm grav}(r)=\left(-\frac{\partial^2}{\partial r^2}-4\pi m \frac{\partial}{\partial r}-4\pi^2 m^2\right)+ mgr~.
\label{p2}
\end{equation}  
The last two terms in the parenthesis  in the above equation represent a deviation from the standard Hamiltonian. This deviation is  a \emph{general} and \emph{unavoidable}  consequence of the entropic description of gravitation and is independent of the details of microscopic physics behind such description. We would like to stress one subtle point here. Although, the Hamiltonian  (\ref{p2}) is non-Hermitian operator, due to the non-Hermiticity of $\hat p$ (\ref{p1}), its eigenvalues are real  \cite{Kobakhidze:2010mn}. This means that time evolution of the system is still unitary. This is, of course, related to the assumed reversibility of the thermodynamical process associated with the entropic force \cite{Verlinde:2010hp}, and hence conservative nature of entropic gravity. Note that conservative entropic and potential forces still differ from each other, since to `undone' the action of entropic force one needs an infinite time, while the action of potential forces can be `undone' in a finite time. Therefore, one needs to ensure that the microscopic dynamics underlying the thermodynamic process is such that the resulting entropic force is indistinguishable from the standard potential forces at least classically. Following \cite{Verlinde:2010hp} we also assume that this is indeed the case.

This is a good place to discuss the criticism of the above formalism expressed in \cite{Chaichian:2011xc}. The authors of \cite{Chaichian:2011xc} question the validity of Eq. (\ref{p0}) mistakenly arguing that it implies that holographic screens at different $r$ have the same number of microstates, and hence their entropy do not depend on $r$. In the view of Eq. (\ref{5}), the total $r$-independence of entropy should not be much of surprise. Indeed, $\frac{dS}{dr}=0$ can be viewed as a defining property of the entropic force, however, the gradient of entropy must, of course, be non-zero, i.e. $\frac{\partial S}{\partial r}\neq 0$. 

Let us consider physics behind  Eq. (\ref{p0}) in more details. The macroscopic system under consideration is two gravitating particles $m$ and $M$ placed at distance $r$ from each other. Microscopically this system can be described by e.g. the holographic screen at $r-\Delta r$  surrounding the particle $M$ and the particle $m$ placed outside the region surrounded by the screen. Screen microstates describe everything in the interior region including particle $M$ and has  no information on the outside region where particle $m$ is placed. Hence, the density matrix of the whole system is a tensor product of the density matrix $\rho_{\cal N}(r)$ associated with the particle $m$ and the screen density matrix $\rho(r-\Delta r)$, $\rho_{\cal N}\otimes \rho(r-\Delta r)$. Removing neutron states correspond to considering the reduced density matrix ${\rm Tr}_{\cal N}\rho_{\cal N}\otimes \rho(r-\Delta r)$ which is nothing but $\rho(r-\Delta r)$. Now, the very same macroscopic system can be described just by the holographic screen at position $r$ with density matrix $\rho(r)$. The particle $m$ is a part of the screen now, and is described by the screen microstates. One must realize, however, that these are two different microscopic descriptions of the same macroscopic system, and therefore the entropy associated with $\rho(r)$ must be equal to the entropy associated with  $\rho_{\cal N}\otimes \rho(r-\Delta r)$. This immediately leads to (\ref{p0}). 

As the counterargument,  it was claimed in  \cite{Chaichian:2011xc} that removing particle $m$ from the screen at $r$ would mean that the resulting density matrix $\rho(r)$ is the same as       
$\rho(r-\Delta r)$, and hence screens at different $r$ have the same entropy. This conclusion is based on the confusion about the `coarse graining' the microstates on the holographic screen. The `coarse graining' of microstates inevitably leads to the change in the macroscopic parameter $r$ such that the screen `moves' towards the particle $M$ \cite{Verlinde:2010hp}.  The coarse grained screen at $r$ is actually coincides with the one at $r-\Delta r$, in full accordance with the physical picture described above. Thus, the conclusion drawn in \cite{Chaichian:2011xc} is  incorrect. 

Another way to see that the conclusion in  \cite{Chaichian:2011xc} concerning the validity of Eq. (\ref{p0}) is false is to recall that the number of microstates at the holographic screen surrounding the particle $M$ is $N(r)=\frac{2M}{T}$ (see Eqs. (3.11) and (3.12) in \cite{Verlinde:2010hp}), while the number of screen microstates associated with particle $m$ is $n(r)=\frac{2m}{T}$ (see Eq. (3.14) in \cite{Verlinde:2010hp}). That is to say, $r$-dependence in both cases is determined through the Unruh temperature, $T=\frac{G_NM}{2\pi r^2}$. As a result $\Delta N(r)/N(r)=\Delta n(r)/n(r)\neq 0$ and we again obtain Eq. (\ref{p0}).

\paragraph{Gravitational bound states of neutron.}

In  remarkable experiments with ultra-cold neutrons in the gravitational field of Earth the first bound state of neutron has been identified \cite{Nesvizhevsky:2002ef} in full agreement with the standard quantum mechanical prediction. The theoretical prediction for neutron bound states within Verlinde's entropic gravity is dramatically different from the standard prediction and is in disagreement with the observations. This case was considered in \cite{Kobakhidze:2010mn}. The solution to the modified Schr\"odinger equation, $\hat H\psi_n=E_n\psi_n,$ can  easily be found:
\begin{equation}
\Psi_n(r)=N_n{\rm e}^{-2\pi\frac{ r}{\lambda}}\psi_n(r)
\label{a1}
\end{equation}
with $E_n=mgr_n+2\pi^2 m$; $N_n$ is the normalization factor, $\lambda=1/m=1.3 \cdot10^{-9} \mu{\rm m}$ is the neutron Compton wavelength and $\psi_n$ is the solution within the standard theory. The experiments demonstrate the existence of the first bound state at $r_1\approx 12.2 \mu{\rm m}$ seen as a step-function raise of intensity of the neutron beam passing through the slit with variable height $h$ at $h=r_1$, as it is predicted by the standard theory.  However, the modified wavefunction (\ref{a1}) is exponentially suppressed relative to the standard solution and, since $\lambda<<<r_1$, the neutron beam would freely propagate through the slit even for $h<<< r_1$. Thus, Verlinde's theory fails to reproduce the observations of the first bound state of neutron in the gravitational field of Earth.  

\paragraph{Neutron interference}

Let us consider now double slit experiment with neutrons.  The observation of gravitationally induced interference has been first reported in \cite{Colella:1975dq}. Again, the observed interference picture is in full agreement with the standard quantum mechanical prediction. On the contrary, Verlinde's entropic approach to gravitation fails to reproduce observations. The qualitative argument why this happens is given in \cite{motl}: difference in microstates of neutron passing two slits destroy the interference picture. This can be readily confirmed within our formalism.

Indeed, using the modified Hamiltonian (\ref{p2}), it is easy to verify that a wave travelling path $r(\lambda)$ acquires an exponential factor relative to the standard solution $\psi_{\lambda}$:
\begin{equation}
\Psi_{\lambda}= {\rm e}^{-2\pi m \int_{r(\lambda)} d\lambda}\psi_{\lambda}~.
\end{equation} 
Therefore, the interference visibility of two waves $\Psi_{\lambda}$ and $\Psi_{\tau}$ will be:
\begin{equation}
V_{\rm Verlinde}\equiv\frac{2|\Psi_{\lambda}||\Psi_{\tau}|}{|\Psi_{\lambda}|^2+|\Psi_{\lambda}|^2}=
\frac{2{\rm e}^{-2\pi m(\int_{r(\lambda)} d\lambda-\int_{r(\tau)} d\tau)}(|\psi_{\lambda}|/|\psi_{\tau}|)}{1+{\rm e}^{-2\pi m(\int_{r(\lambda)} d\lambda-\int_{r(\tau)} d\tau)}(|\psi_{\lambda}|/|\psi_{\tau}|)^2}~.
\end{equation}
 Assuming that two slits are at $r$ and $r+\Delta r$ ($\Delta r > 0$) distances from the Earth we find that the interference visibility  $V_{\rm Verlinde}$  within the Verlinde's theory is:
 \begin{equation}
 V_{\rm Verlinde}\approx {\rm e}^{-2\pi\frac{\Delta r}{\lambda}}~,
 \end{equation}
 for the ideal visibility $V_{\rm standard}=1$ within the standard treatment. Since in typical experiments the distance $\Delta r$ is many orders of magnitude bigger than the neutron Compton wavelength $\lambda$, we find that $V_{\rm Verlinde}\approx 0$. Thus, Verlinde's theory implies practically complete decoherence of neutron waves and thus non-observation of the gravitationally induced interference pattern.  This strikingly  contradicts observations.

\paragraph{Conclusion.} In this paper, we have reconfirmed our previous conclusion \cite{Kobakhidze:2010mn} that Verlinde's entropic gravity when applied to quantum-mechanical systems shows large deviation from the standard approach. We have also demonstrated  that the  opposite conclusion reached by the authors of \cite{Chaichian:2011xc} is erroneous.  Our conclusion is independent of the fine details of Verlinde's theory and steams from the very basic assumption about the entropic nature of gravitation and inertia. Both classic experiments   \cite{Nesvizhevsky:2002ef} and \cite{Colella:1975dq} are in sharp contradiction with the predictions of Verlinde's theory. Thus,  we restate: gravity is not an entropic force.

\paragraph{Acknowledgements:} 
I would like to thank the authors of \cite{Chaichian:2011xc} for providing their manuscript before its publication and for the email exchange concerning disputed issues. This work was supported in part by the Australian Research Council.


\end{document}